# An Infrared Communication System based on Handstand Pendulum

Xingchen Li[1], Changlu Li[2], Yun Wang[1] and Mengqi Lei[1]

(1. School of Microelectronics, Tianjin University, Tianjin, 300072, China;

2. School of Electrical and Information Engineering, Tianjin University, Tianjin, 300072, China)

**Abstract**    This paper mainly introduces an infrared optical communication system based on stable and handstand pendulum. This system adopts the method of loading the infrared light emitting end on an handstand pendulum to realize the stability and controllability of the infrared light transmission light path. In this system, 940nm infrared light is mainly used for audio signal transmission, and an handstand pendulum based on PID is used to control the angle and stability of infrared light emission. Experimental results show that the system can effectively ensure the stability of the transmission optical path and is suitable for accurate and stable signal transmission in bumpy environments.

**Key words**    PID Algorithm, Infrared Communication, Band Pass Filter, Power Amplifier, Automotive Control

## Introduction

With the development of modern communication technology, optical communication has become an indispensable communication method. Among them, the mode of using infrared light for communication is a potential development direction. At present, a large number of researchers are studying the infrared optical communication neighborhood in China.

Infrared optical communication is a communication method that uses electromagnetic waves in the infrared light segment for signal transmission [1]. Infrared has the characteristics of large capacity, strong confidentiality, and good anti-electromagnetic interference performance, which makes it have the potential to be applied to communication systems. In the existing infrared optical communication system, the transmitting and receiving end usually use infrared pair tubes, so that in the analog channel In transmission, since the pair of tubes can be equivalent to a transistor, the process of modulation and demodulation can be omitted [2]. In view of some of the original infrared communication problems, there have been many research results in recent years, which have prominently solved the problems of low signal quality and poor controllability. For example, the existing research results mainly include: direct modulation of analog voice signals to infrared Channel [3], adding a filter circuit to improve signal quality [4], modulating the voice signal into a digital signal, and then using infrared channel transmission [5], and selecting a single-chip microcomputer [6], FPGA to improve system performance, etc. The transmission quality of infrared light signals thus has been improved to a certain extent.

However, for certain dynamic environments, such as a sea-going vessel communicating in a strong turbulence environment, infrared optical communication will face the problem of being unable to receive stable direct radiation. In addition, there is currently no way to precisely adjust the direction of the infrared light beam, and the current alignment still mainly uses manual alignment. These problems have brought great challenges to infrared optical communications, especially the stability and reliability of infrared optical transceivers.

Based on these problems, this paper proposes an infrared optical communication system based on an inverted pendulum. An inverted pendulum is a device that uses a control algorithm to achieve a pendulum bar inverted. It can load various devices that need to work normally under stable conditions on the pendulum bar to achieve stable balance and other effects [7]. In this paper, the inverted pendulum platform is equipped with an infrared light emitting device, so that the emitted light beam is not affected by the bumpy environment, so as to realize the infrared light transceiver device to achieve a highly directional and stable infrared light communication system. In addition, the system can also adjust the angle of the emitted infrared light beam, making the alignment more convenient.

This article mainly consists of the following parts. The first part introduces the principle and scheme of the system, mainly introduces the design principles of each part and the main parameters considered; the second part introduces the hardware construction of the system, specifically introduces the hardware composition of the infrared light system and the inverted pendulum system; In the third part, performance of the system is evaluated and verified.

# 1 System principle and scheme design

The infrared optical communication system introduced in this article is an integrated integration of infrared optical communication subsystem and self-stabilizing inverted pendulum subsystem. This part will combine the system block diagram to introduce the basic principles and implementation schemes of the infrared optical communication part and the inverted pendulum part.

1.1 Principle of infrared optical communication system

The basic block diagram of the infrared optical communication system is shown in Figure 1.1. The modules mainly used in this infrared light system include audio small signal amplification, infrared emission, infrared reception, low-pass filtering and power amplification. In this paper, the audio signal is mainly used as the transmitted signal for research.

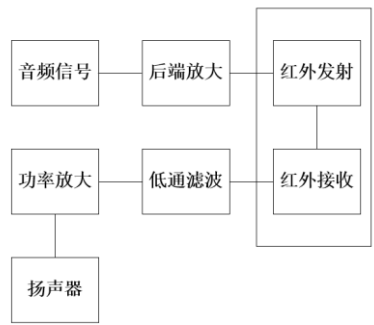

**Figure 1.1. Block diagram of infrared optical communication system**

The following will specifically introduce the basic principles involved in the infrared transmitting and receiving part. So far, infrared light transceiver can be mainly divided into analog channel and digital channel. When infrared optical communication uses a digital channel, it is necessary to convert the signal into a digital quantity after sampling, and use the on-off transmission of the coupling tube [4]. The main advantages of this method are: the signal is relatively stable and it is not easy to introduce noise [5]. However, the shortcomings of using digital channel transmission, such as aliasing caused by improper sampling rate settings, need to cooperate with digital-to-analog-to-digital conversion equipment, which leads to high costs, which cannot be avoided. Therefore, this article does not adopt this scheme.

This text uses analog quantity to carry on the signal transmission, rely on the signal strength change to make the infrared light emitting tube light intensity change synchronously to transmit the signal. This method is easy to be affected by the external environment, but it still has a good effect in indoor short-distance communication [12], and the circuit structure is simple (no modulation and demodulation circuit is required), so this scheme intends to use the analog signal transmission method.

Common communication systems need to involve modulation and demodulation in the process of sending and receiving signals. Taking the amplitude modulation signal as an example, for the fundamental wave signal f(t), when the high frequency signal $A\sin\omega t$ is modulated, the carrier signal formed is:

$$f_{out}(t) = f(t)A\sin\omega t \qquad (1.1)$$

However, if a reasonable transceiver device (such as an infrared light coupling tube) is used, the modulation and demodulation process at the transceiver end can be avoided. Specifically, when infrared optical communication uses optical signals to transmit information, since the light intensity changes with the current passing through the infrared light emitting diode, and the resistance of the receiving diode changes with the received light intensity, it is only necessary to convert the fundamental wave signal into The current signal can use the coupling of the photoelectric transmitting tube and the receiving tube to convert the current signal of the transmitting end into the resistance signal of the receiving end, and then into the current signal of the receiving end [13]. In this process, since the light frequency is fixed and the light intensity changes with the current, the amplitude modulation and demodulation process is automatically completed, so there is no need to design additional complex modulation and demodulation circuits. This article uses this scheme for time control. In this case, the system output signal can be expressed by the following formula:

$$f_{out}(t) = f(t)A(t) \quad (1.2)$$

Among them, fout(t) is the output current signal, f(t) is the input audio signal, and A(t) is the equivalent modulation and demodulation function of the transceiver device. In the infrared light coupling tube, A(t) is equal to the forward transconductance of the coupling tube system.

1.2 Self-stabilizing inverted pendulum system

In this paper, an inverted pendulum system is used to align and calibrate the optical path of infrared light transmission. Specifically, load the infrared transmitter module shown in Figure 1.1 on the inverted pendulum to control parameters such as the launch angle of the transmitter front end. This section will specifically introduce the principle of the inverted pendulum system.

The inverted pendulum system is a typical multi-variable, high-order, non-linear, strong coupling, natural unstable system, which is widely used in adaptive control, variable structure control, fuzzy logic control and other fields [8]. The usual inverted pendulum system realizes control by calculating the angle as the feedback quantity. The basic model of the inverted pendulum system used in this article can be simplified as shown in Figure 1.2.

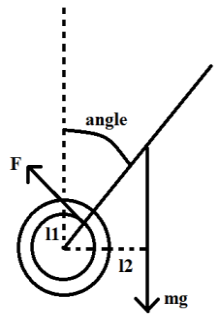

**Figure 1.2. Inverted pendulum model**

The equations used to describe the system will be calculated below. In this system, the feedback quantity is the angle θ between the inverted pendulum and the vertical direction (the set angle, that is, the angle when the pendulum is expected to be stable), and the output quantity is the force F that the motor acts on the inverted pendulum. In this mechanical structure, there are mainly two forces acting on the pendulum, namely the gravity of the pendulum and the tangential traction force of the motor on the pendulum. From the angular momentum theorem, the following formula can be obtained:

$$\frac{\partial L}{\partial t} = M = F \times l1 - mg \times l2 \quad (1.3)$$

Among them, the angular momentum of the single rod shaft at one end can be expressed as:

$$L = I\omega = \frac{ml^2}{3}\omega \quad (1.4)$$

l is the length of a single pole. Combining equations (1.3) and (1.4) together are:

$$\frac{\partial L}{\partial t} = \frac{ml^2}{3}\frac{\partial \omega}{\partial t} = \frac{ml^2}{3}\beta$$

$$= F \times l1 - mg \times l2 = F \times l1 - mg\frac{l}{2}\sin\theta \quad (1.5)$$

From the relationship between angle and angular acceleration:

$$\beta = \frac{\partial^2 \theta}{\partial t^2} \quad (1.6)$$

Therefore, when the angle is very small, the following differential equation can be obtained:

$$\frac{ml^2}{3}\frac{\partial^2 \theta}{\partial t^2}$$

$$= F \times l1 - mg\frac{l}{2}\sin\theta = F \times l1 - mg\frac{l}{2}\theta \quad (1.7)$$

This is the system equation of the inverted pendulum. The block diagram of the inverted pendulum system used in this article is shown in Figure 1.3.

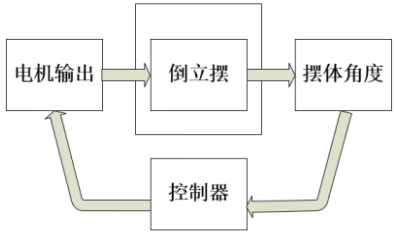

**Figure 1.3. Inverted pendulum system block diagram**

1.3 Inverted pendulum control algorithm principle

This article intends to use PID algorithm to control the inverted pendulum system. Based on (1.7), this part will mainly introduce the principle and design of the control algorithm.

The PID algorithm is controlled by proportional, integral, and differential. Its advantages include simple principle, convenient use, strong adaptability, and strong robustness. In fact, in the process of realizing the inverted pendulum, since the motor is used as a servo system, there is no steady-state error, so you can choose to use only PD control [10]. Do laplace transformation on (1.7), and get:

$$\frac{ml^2}{3}s^2\Theta(s) = F(s)l_1 - mg\frac{l}{2}\Theta(s) \quad (2.7)$$

The two main poles of the solution are:

$$s = \pm j\sqrt{\frac{3g}{2l}} \quad (2.8)$$

This is the extreme case without feedback. The two poles fall on the jw axis and oscillate with equal amplitude, so the system cannot be stabilized. Add a pair of proportional quantities (Kp) and differential quantities (Kd) to the system, and let F be the external force acting on the pendulum, then:

$$\frac{ml^2}{3}s^2\Theta(s) = F(s)l_1 - K_p\Theta(s) - K_d s\Theta(s) - mg\frac{l}{2}\Theta(s) \quad (2.9)$$

Then the system transfer function can be obtained as:

$$H(s) = \frac{l_1}{\frac{ml^2}{3}s^2 + K_d s + K_p + mg\frac{l}{2}} \quad (2.10)$$

This formula shows that in this system, as long as:

$$\begin{cases} K_d^2 - \frac{4ml^2}{3}\left(K_p + mg\frac{l}{2}\right) < 0 \\ K_d > 0 \end{cases} \quad (2.11)$$

Then the two main poles of the system can be located at the left end of the jw axis. And under this condition, the larger the Kd, the more stable the system. Based on this, the pseudo code implemented by the PID algorithm is shown in 1.4.

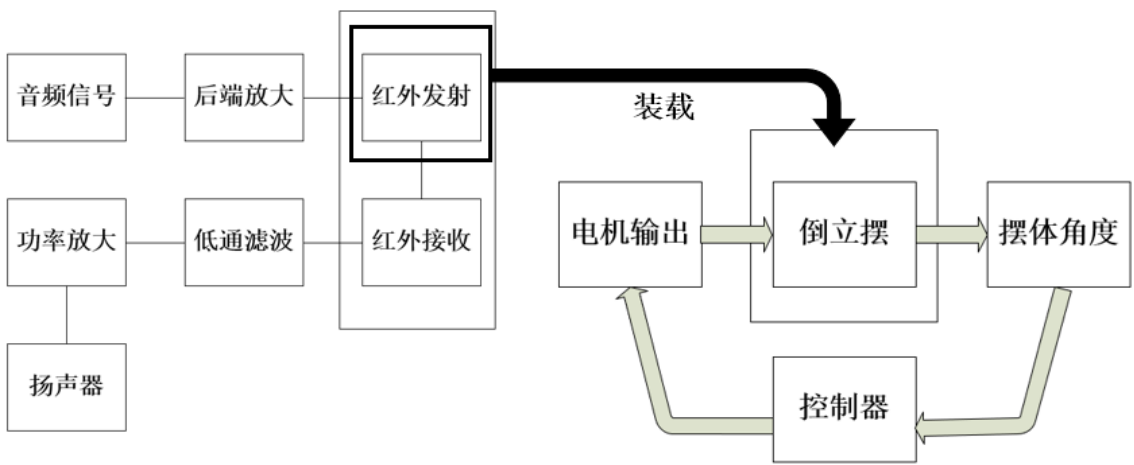

**Figure 1.5. System block diagram**

```
last_angle = new_angle;
new_angle = input_angle;
Error_angle = new_angle-last_angle;
D_error_angle = (new_angle-last_angle)/dt;
F_out = -Kp*Error_angle-Kd* D_error_angle /10;
```

**Figure 1.4. Inverted pendulum system control pseudo code**

## 2 System hardware implementation

This part will mainly introduce the hardware realization of the above infrared optical communication system and inverted pendulum system. The overall system block diagram is shown in Figure 1.5:

### 2.1 Infrared communication system design

The infrared transmitting and receiving circuit of this scheme is mainly composed of high-frequency small-signal audio amplifier composed of NE5532AI, infrared transmitting circuit, infrared receiving circuit, and power amplifier circuit composed of TDA2822M; at the same time, NE5532AI is used as a band during the design of the scheme to be the active part of the pass filter.

2.1.1 Signal acquisition and back-end amplification

For the sound signal to be transmitted, this article uses a four-wire control headphone cable to collect, and only uses one of the channels. In the audio amplification part, this article uses a basic voltage amplification method. Let the op amp work in the deep negative feedback area, using the principles of "virtual short" and "virtual break", the circuit topology designed in this article is shown in Figure 2.1:

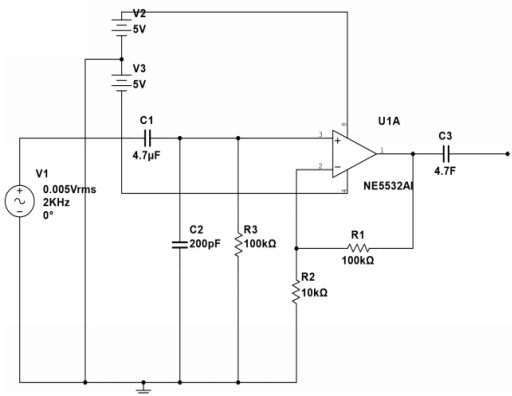

**Figure 2.1. Signal amplifier circuit**

According to the components and circuit connection methods in the figure, the voltage gain A of the circuit is calculated to satisfy:

$$A_V = 1 + \frac{R_1}{R_2} \qquad (3.3)$$

2.1.2 Infrared transceiver

This text uses infrared light coupling tube to realize the infrared light transmission and reception of audio signal.

The usual scheme uses the typical circuit shown in Figure 2.2 to realize signal transmission, in which the triode plays the role of adjusting the input signal amplitude and improving the sound quality. In some schemes, the triode can be replaced by the integrated amplifier chip.

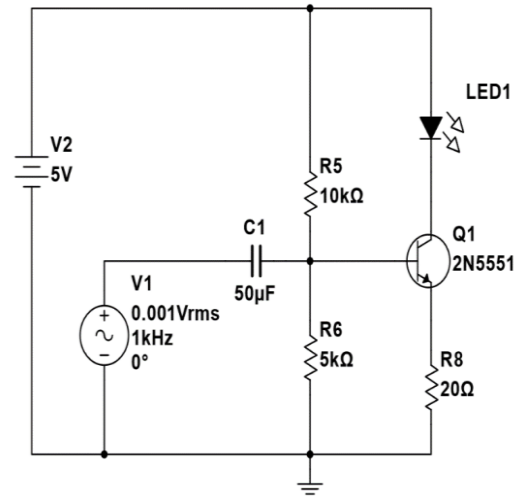

**Figure 2.2. Typical triode infrared transmitting circuit**

However, when using this method for transmission, the use of transistors and other devices for bias and amplification, because the relationship between Vbe and Id is:

$$I_d = \beta e^{(\frac{V_{be}-V_{th}}{V_T})} \qquad (3.5)$$

Therefore, an exponential harmonic is inevitably introduced, which is equivalent to causing interference to the original signal. This problem can be avoided by using direct resistance bias as shown in Figure 2.3 in this article.

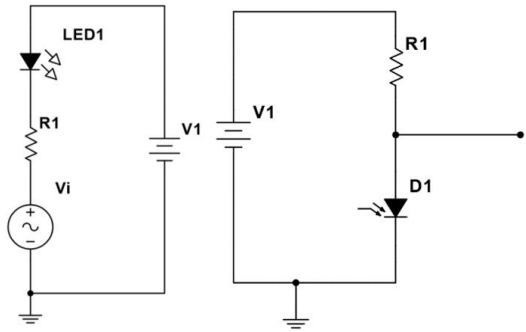

**Figure 2.3. Resistance bias infrared transceiver circuit**

The voltage applied to the LED is the superposition of the power supply voltage and the input audio signal voltage:

$$V_{LED1} = V_1 - V_i \qquad (3.4)$$

Therefore, the voltage at both ends of the LED well reflects the change of the signal voltage, and the function of modulating the infrared light into an amplitude modulation wave carrying the audio signal is realized.

### 2.1.3 Filter circuit design

In order to filter out the noise introduced in the transmission process, a band-pass filter is needed to filter the signal. FFT actual measurement of the pre-stage circuit, it is found that the pre-stage signal is mainly concentrated in 100Hz-3KHz, and the noise signal mainly includes the background noise distributed in the entire frequency spectrum, and some low-frequency (<10Hz) and high-frequency (>10kHz) Weight. Based on this, this paper designs the following fourth-order low-pass and third-order high-pass cascaded band-pass filter circuits (Figure 2.5).

Two second-order low-pass cascades are used at the input, mainly to increase the response roll-off coefficient and reduce the upper limit frequency; using third-order high-pass, in addition to making the response steep, it can also effectively avoid near the lower limit frequency Of jitter.

In order to make the design meet the frequency requirements of this article,

In the selection of the resistance and capacitance of the type, use the formula:

$$k = \frac{100}{f_c C} \qquad (3.6)$$

The final resistance and capacitance values are shown in Table 1:

**Table 1 Circuit component value parameter table**

| Low Pass | C1=5nF | C2=15nF |
|---|---|---|
| | R1=6.8K | R2=3.6K |
| High Pass | C3=1uF | C4=1uF |
| | R3=5K | R4=10K |
| | C5=1uF | R5=10.4K |

### 2.1.4 Power amplification

The sound signal at the receiving end can drive the speaker to emit sound only after the audio power is amplified. This system uses TDA2822 audio power amplifier chip, and the designed circuit is as follows:

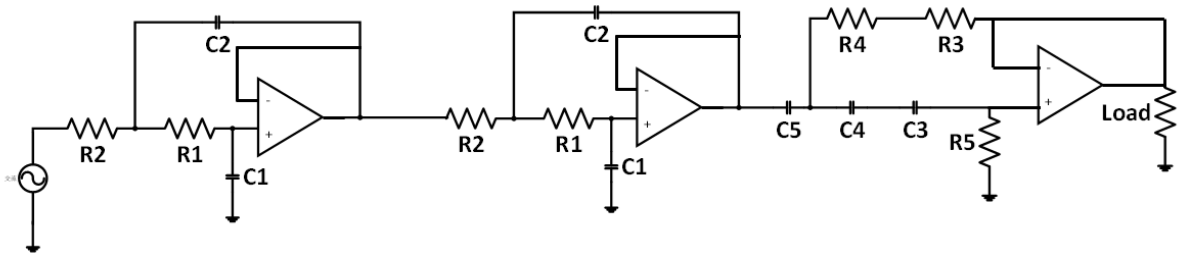

**Figure 2.4. Band pass filter circuit**

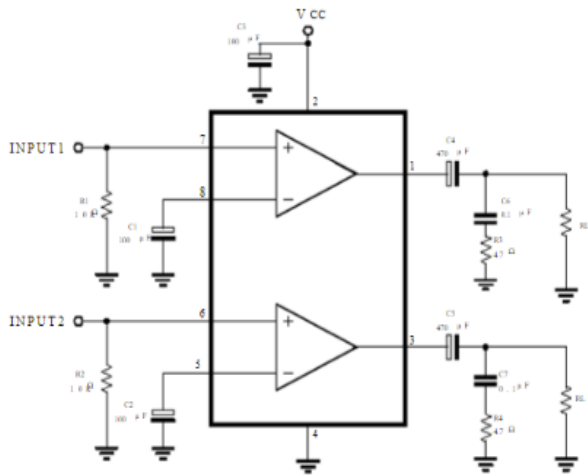

**Figure 2.5. Power amplifier circuit**

## 2.2 Self-stabilizing inverted pendulum system design

The frame part of the inverted pendulum is mainly composed of a control circuit and a controller. Circuit design part. In order to realize the inverted pendulum, the main circuits include the system circuit of the controller, the drive circuit, the signal transmission circuit and the circuit connected with the peripherals.

In this system, the drive circuit adopts the H full-bridge drive circuit, which is controlled by the IR2104 chip, which can ensure that the corresponding voltage can be formed between the bridge arms under the condition of PWM drive. The relationship between PWM duty cycle duty and output voltage is as follows:

$$V_{out}(duty) = V_{ref} \times duty \quad (2.6)$$

The circuit diagram of the driving part is shown in 2.6(a). The thing diagram of the driving circuit is shown in 2.6(b).

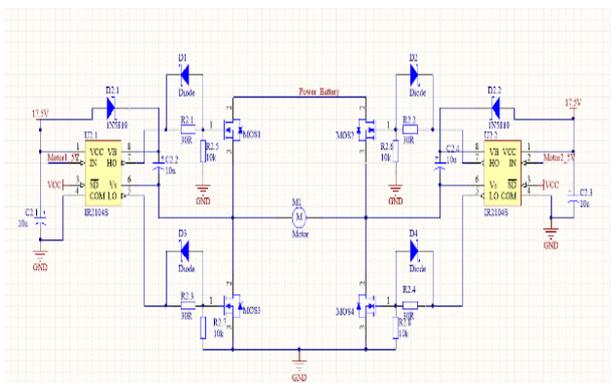

(a)

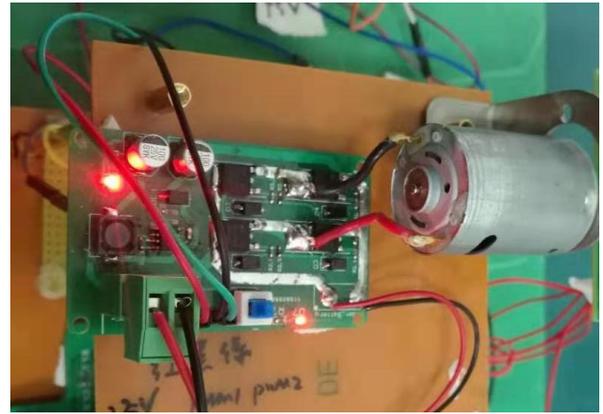

(b)

**Figure 2.6. (a) H full bridge drive circuit (a) H full bridge drive circuit board**

The system control and transmission circuit rely on K60DN512Z chip to cooperate with the peripheral circuit to realize.

The method of loading the infrared light emitting end to the inverted pendulum is introduced below. In this inverted pendulum system, as shown in Figure 2.7, the circuit board where the infrared light emitting end is located is directly latched on the board frame of the inverted pendulum by copper column screws, thus ensuring the absolute connection between the two.

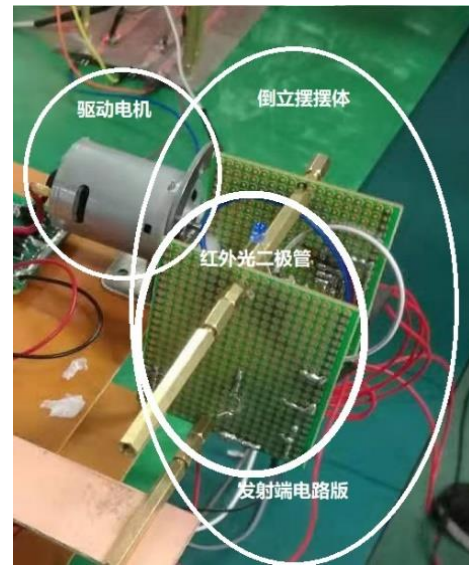

**Figure 2.7. Connection of infrared light emitting end and inverted pendulum**

## 3 System assembly and testing

This part mainly introduces the assembly, test and main experimental results of the system.

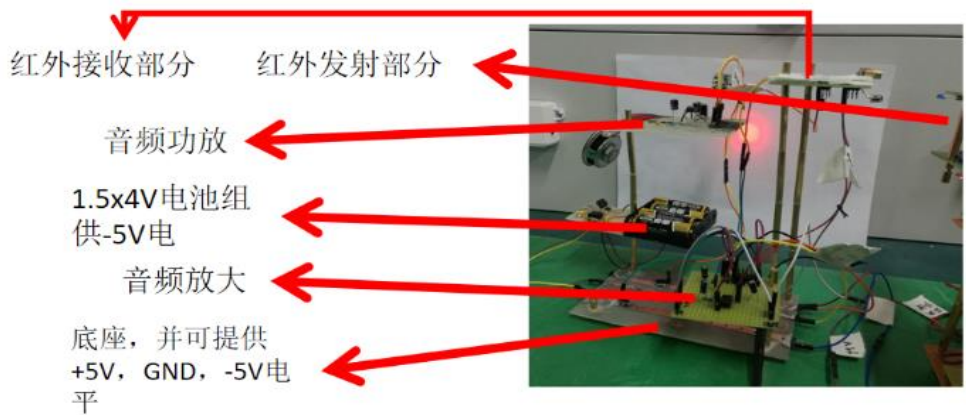

(a)

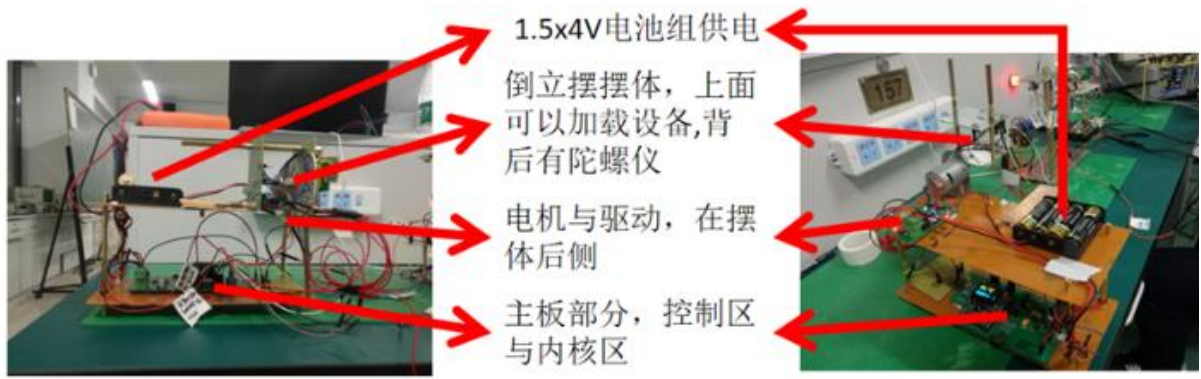

(b)

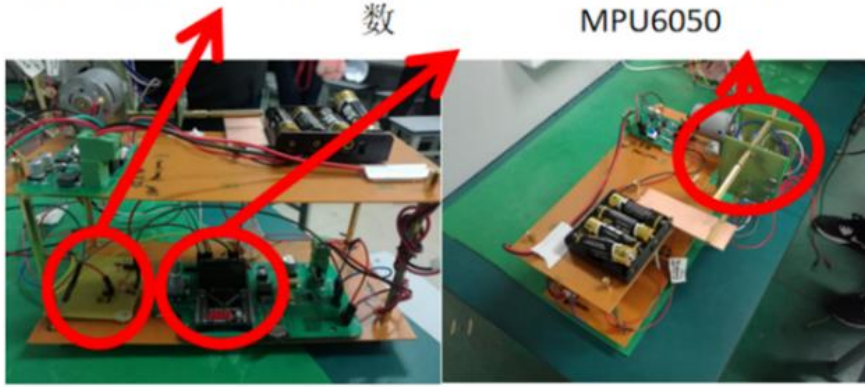

（c）

Figure 3.1. Assembly drawing of infrared light system

## 3.1 System assembly

In the system assembly, it is mainly necessary to consider the interconnection of the components of the infrared light system, the interconnection of the components of the inverted pendulum system and the coupling between the infrared light and the inverted pendulum system. Figure 3.1 shows the assembly results of each part.

First introduce the assembly of the infrared light system. The final assembly platform used in the test is shown in Figure 3.1(a). Among them, each circuit of this part is made of independent circuit boards and assembled into the structure shown in the figure by using copper pillars. Such a structure makes the most of the advantages of height in the space, making the space utilization rate higher at the vertical level.

The assembly of the mechanical structure of the inverted pendulum is shown in Figure 3.1(b). The mechanical structure assembly adopts the upper and lower two-layer structure, in which the control circuit and so on are placed on the bottom layer to facilitate the operation of the system, and the inverted pendulum drive motor, battery pack, etc. are placed on the top layer, so that the inverted pendulum has enough in the actual swing process space.

The assembly of the inverted pendulum control circuit and pendulum body is shown in Figure 3.1(c). There is an OLED display on the main control circuit that can display the actual angle and set angle of the pendulum in real time, and the knob on the circuit board can adjust the set angle, Kp, Kd and other parameters. The inverted pendulum body is loaded with the infrared light emitting part, and the specific loading method has been introduced earlier.

### 3.2 System test

The test of the system mainly includes the test of the stability of the inverted pendulum, the test of the infrared optical channel transmission and the test of the stability of the infrared light transmission based on the inverted pendulum.

In terms of the performance of the inverted pendulum, it mainly focuses on the actual stable angle after loading the infrared light emitting end under different set angles. The stable angle is measured by a level gauge, and the test results are shown in Table 2. It can be seen that when the absolute value of the set angle is less than 15°, the inverted pendulum can guarantee an angle with a small difference from the set angle after being stabilized. In fact, in real life, it is difficult for the infrared optical communication system application site to have large bumps above 15°, so the design meets the requirements.

In terms of infrared light performance, it mainly investigates the signal situation at the output end of the power amplifier at the receiving end through the infrared light transceiver system when a 1kHz sine signal is input. Figure 3.2 shows the normalized (namely amplitude/maximum amplitude) signal waveform, where the blue signal is the input signal and the red signal is the output signal. It can be seen that the output waveform has a delay of about 0.15ms compared to the input waveform, and a certain harmonic is generated. However, as a whole, the 1kHz oscillation signal is still the main signal composition. Therefore, within the allowable range of error, the signal meets the requirements of transmission accuracy.

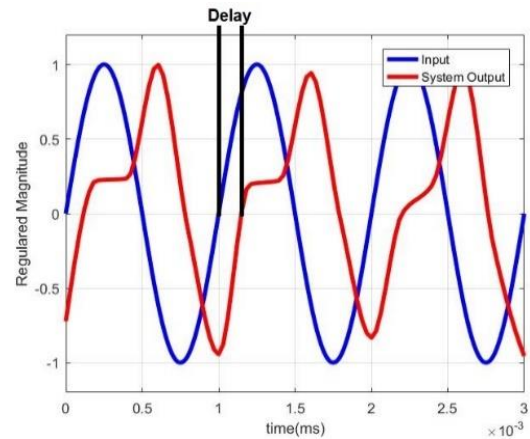

**Figure 3.2. Waveform diagram of infrared optical communication system**

For the test of the stability of infrared light

**Table 2 Inverted pendulum function test**

| Theoretical stable angle (degrees) | -20 | -15 | -10 | 0 | 10 | -15 | -20 |
|---|---|---|---|---|---|---|---|
| Actual stable angle (degrees) | Large oscillation | Small oscillation | -11.5 | 1 | 11 | Small oscillation | Large oscillation |

transmission, we placed the system on a movable desktop, and loaded it on the inverted pendulum at the transmitter and directly connected the transmitter (that is, the transmitter was firmly connected to the base, and the output direction could not be maintained. In the case of changing), a complete and continuous sound signal is input, and the continuity and integrity of the input sound signal are judged. The results show (Table 3), when the inverted pendulum is loaded, there is almost no intermittent sound. Therefore, the system can ensure the stability of infrared light transmission in a bumpy environment

Table 3 System test results

|  | Turn on | Turn off |
|---|---|---|
| **Stable** | Complete audio | Complete voice signal |
| **Vertical vibration** | Complete audio | Intermittent audio |
| **Parallel front-back vibration** | Complete audio with noise | Audio vanish |
| **Parallel left-right vibration** | Intermittent audio | Audio vanish |

## 4 Conclusion

In this paper, an infrared optical communication system based on an inverted pendulum can be used in an ideal place to achieve stable signal transmission using 940nm infrared light analog transmission under the condition that the optical transmission direction is fixed. Among them, the infrared light system can better transmit audio signals below 6KHz, and the inverted pendulum system can be reset to ensure that the infrared light emission direction remains unchanged when the offset is less than 15°, instead of manually adjusting the infrared light emission accurately direction. Better stability and transmission accuracy have certain application significance.

## Acknowledgement

Thank you for the establishment of the electronic technology practice course of Tianjin University and the electronic and electrical experimental teaching center for the help of this article in the instrument, equipment and experimental environment. The work has been completely implemented in Oct, 2018.